\documentclass[aps,pra,reprint,superscriptaddress,longbibliography,
               nofootinbib,amsmath,amssymb,floatfix]{revtex4-2}
\usepackage{graphicx}
\usepackage{bm}
\usepackage[hidelinks]{hyperref}

\makeatletter
\let\VfiveDefaultFootnote\frontmatter@thefootnote
\renewcommand{\frontmatter@thefootnote}{%
  \ifcase\value{\@mpfn}\or\ensuremath{\dagger}\or\ensuremath{*}%
  \else\VfiveDefaultFootnote\fi
}
\AtEndDocument{%
  \if@filesw
    \immediate\write\@auxout{\string\citation{v5ReferencePages}}%
  \fi
}
\makeatother

\begin{document}

\title{Collective coupling of over 100 individually trapped atoms to a single cavity mode}

\newcommand{\EqualContribution}{These authors contributed equally to this work.}
\newcommand{\CorrespondenceNotes}{%
  Correspondence\protect\newline%
  Zhihui Wang \quad wangzhihui7150@sxu.edu.cn\protect\newline%
  Gang Li \quad gangli@sxu.edu.cn\protect\newline%
  Tiancai Zhang \quad tczhang@sxu.edu.cn%
}

\author{Shijun Guan}
\thanks{\EqualContribution}
\affiliation{State Key Laboratory of Quantum Optics Technologies and Devices and Institute of Opto-Electronics, Shanxi University, Taiyuan, 030006, Shanxi, China.}
\affiliation{Collaborative Innovation Center of Extreme Optics, Shanxi University, Taiyuan, 030006, Shanxi, China.}

\author{Guansheng Teng}
\thanks{\EqualContribution}
\affiliation{State Key Laboratory of Quantum Optics Technologies and Devices and Institute of Opto-Electronics, Shanxi University, Taiyuan, 030006, Shanxi, China.}
\affiliation{Collaborative Innovation Center of Extreme Optics, Shanxi University, Taiyuan, 030006, Shanxi, China.}

\author{Zhiguo Huang}
\thanks{\EqualContribution}
\affiliation{China Mobile (Suzhou) Software Technology Company Limited, Suzhou 215163, China.}
\affiliation{China Mobile (Hong Kong) Innovation and Research Institute, Hong Kong, China.}

\author{Zhihui Wang}
\thanks{\CorrespondenceNotes}
\affiliation{State Key Laboratory of Quantum Optics Technologies and Devices and Institute of Opto-Electronics, Shanxi University, Taiyuan, 030006, Shanxi, China.}
\affiliation{Collaborative Innovation Center of Extreme Optics, Shanxi University, Taiyuan, 030006, Shanxi, China.}

\author{Pengfei Yang}
\affiliation{State Key Laboratory of Quantum Optics Technologies and Devices and Institute of Opto-Electronics, Shanxi University, Taiyuan, 030006, Shanxi, China.}
\affiliation{Collaborative Innovation Center of Extreme Optics, Shanxi University, Taiyuan, 030006, Shanxi, China.}
\affiliation{Hefei Qianyuan Quantum Technology Co., Ltd., Hefei, 230088, Anhui, China.}

\author{Pengfei Zhang}
\affiliation{State Key Laboratory of Quantum Optics Technologies and Devices and Institute of Opto-Electronics, Shanxi University, Taiyuan, 030006, Shanxi, China.}
\affiliation{Collaborative Innovation Center of Extreme Optics, Shanxi University, Taiyuan, 030006, Shanxi, China.}

\author{Gang Li}
\thanks{\CorrespondenceNotes}
\affiliation{State Key Laboratory of Quantum Optics Technologies and Devices and Institute of Opto-Electronics, Shanxi University, Taiyuan, 030006, Shanxi, China.}
\affiliation{Collaborative Innovation Center of Extreme Optics, Shanxi University, Taiyuan, 030006, Shanxi, China.}
\affiliation{Hefei Qianyuan Quantum Technology Co., Ltd., Hefei, 230088, Anhui, China.}

\author{Tiancai Zhang}
\thanks{\CorrespondenceNotes}
\affiliation{State Key Laboratory of Quantum Optics Technologies and Devices and Institute of Opto-Electronics, Shanxi University, Taiyuan, 030006, Shanxi, China.}
\affiliation{Collaborative Innovation Center of Extreme Optics, Shanxi University, Taiyuan, 030006, Shanxi, China.}

\begin{abstract}
Coupling optical tweezer arrays to high-finesse cavities provides a shared optical interface for
neutral-atom registers. Scaling this interface to large tweezer arrays requires placing many traps
within the coupling region of a single cavity mode while maintaining uniform trapping and
site-resolved detection. Here we match a two-dimensional array of 192 optical tweezers to the
coupling profile of a millimeter-scale Fabry--Pérot cavity. To improve trap uniformity in the array
generated by crossed acousto-optic deflectors, we combine optimization with optical feedback and
atomic fluorescence feedback, achieving a mean cesium loading probability of approximately 58\%. In
each trial, fluorescence imaging determines whether each individual tweezer site is occupied,
allowing the cavity transmission spectra to be grouped by atom number. The grouped spectra exhibit
resolved normal-mode doublets, and the group with 120 imaged atoms yields an effective collective
coupling of \(G_{\mathrm{\mathrm{eff}}}/2\pi = 21.30 \pm 0.23\) MHz. A model based on the
calculated coupling profile and measured site loading probabilities predicts that the mean squared
coupling per occupied site changes little between 97 and 120 atoms. The fitted collective coupling
is therefore consistent with square-root scaling with atom number despite the spatial variation in
single-atom coupling. These results provides an experimental basis for studying photon-mediated
many-body dynamics and developing cavity-assisted readout and optical links between neutral-atom
registers.
\end{abstract}

\maketitle

\section{Introduction}
\label{sec:introduction}

Two-dimensional optical tweezer arrays provide a versatile platform for neutral-atom quantum
computing~\cite{evered2023gates,chung2025logical,bluvstein2026architecture} and quantum
simulation~\cite{ebadi2021phases,scholl2021antiferromagnets,manovitz2025coarsening} by combining
scalable trapping geometries with spatially resolved access to individual
atoms~\cite{barredo2016assembler,endres2016assembly,sheng2021preparation,lin2025assembly,li2025fiber,manetsch2025array}.
Optical cavities offer complementary capabilities by enhancing atom--photon
interactions~\cite{thompson1992splitting,oh2024coherence} and interfacing atomic states with
propagating optical fields~\cite{ritter2012network,daiss2021gate,ghosh2021photons}. Integrating the
two platforms brings individually trapped atoms into a common optical mode while preserving spatial
access to the array. The shared mode can support cavity-assisted readout and photon-mediated
interactions within an atomic register~\cite{ding2016memories}, as well as optical links between
registers in modular quantum processors~\cite{wachter2019microcavities,shaw2026microscope}.

Experiments with intracavity arrays have demonstrated cavity-assisted mid-circuit measurements,
collective coupling and geometry-dependent collective
scattering~\cite{deist2022measurement,liu2023coupling,yan2023scattering}. The assembly of
intracavity atomic registers and the subwavelength positioning of individual atoms with optical
tweezers have highlighted the role of spatial control in cavity-based quantum network
nodes~\cite{hartung2024register,seubert2025positioning}. Related experiments have coupled
defect-free arrays strongly to optical cavities~\cite{wang2025cavity} and combined cavity coupling
with Rydberg excitation~\cite{desantis2026rydberg}. Extending collective cavity coupling to arrays
of more than 100 individually trapped atoms requires placing many traps within the coupling region
of a single mode while maintaining reliable trapping and site-resolved detection. As the array
spans more of the mode profile, the collective optical response depends on both how many atoms are
present and which sites they occupy.

Maintaining uniform trap intensities is a further challenge when scaling intracavity tweezer arrays
generated by crossed acousto-optic deflectors (AODs). Previous studies have improved array
uniformity through multitone waveform optimization, suppression of intermodulation, and feedback
based on atomic
loading~\cite{wang2023intermodulation,sun2024optimization,mittenbuehler2025synthesis,schymik2022loading}.
Successive diffraction by the two deflectors adds a constraint on the choice of radio-frequency
(RF) tone spacings. Distinct pairs of tones can impart the same net optical frequency shift to
beams directed to different sites, allowing those beams to interfere where their fields
overlap~\cite{decruyenaere2026projections}. The axial and radial tone spacings must therefore be
chosen together, taking into account both the array footprint within the cavity coupling region and
the optical frequency separations between the diffracted beams.

Here we investigate collective coupling at the hundred-atom scale. We match the geometry of a
\(32\times6\) tweezer array to the coupling profile of a millimeter-scale Fabry--Pérot cavity. And
we choose RF tone spacings to avoid exact frequency degeneracy among the target first-order beams.
We combine optical optimization with camera and atomic feedback to improve array uniformity.
Single-atom spectroscopy maps the spatial variation in coupling and guides alignment. In each
trial, fluorescence imaging identifies the occupied sites and provides the atom count used to group
the cavity transmission measurements. We then use the calculated coupling profile and measured site
loading probabilities to examine how spatial variations in single-atom coupling affect the expected
scaling with atom number.

\begin{figure*}[!tbp]
  \centering
  \includegraphics[width=0.92\textwidth]{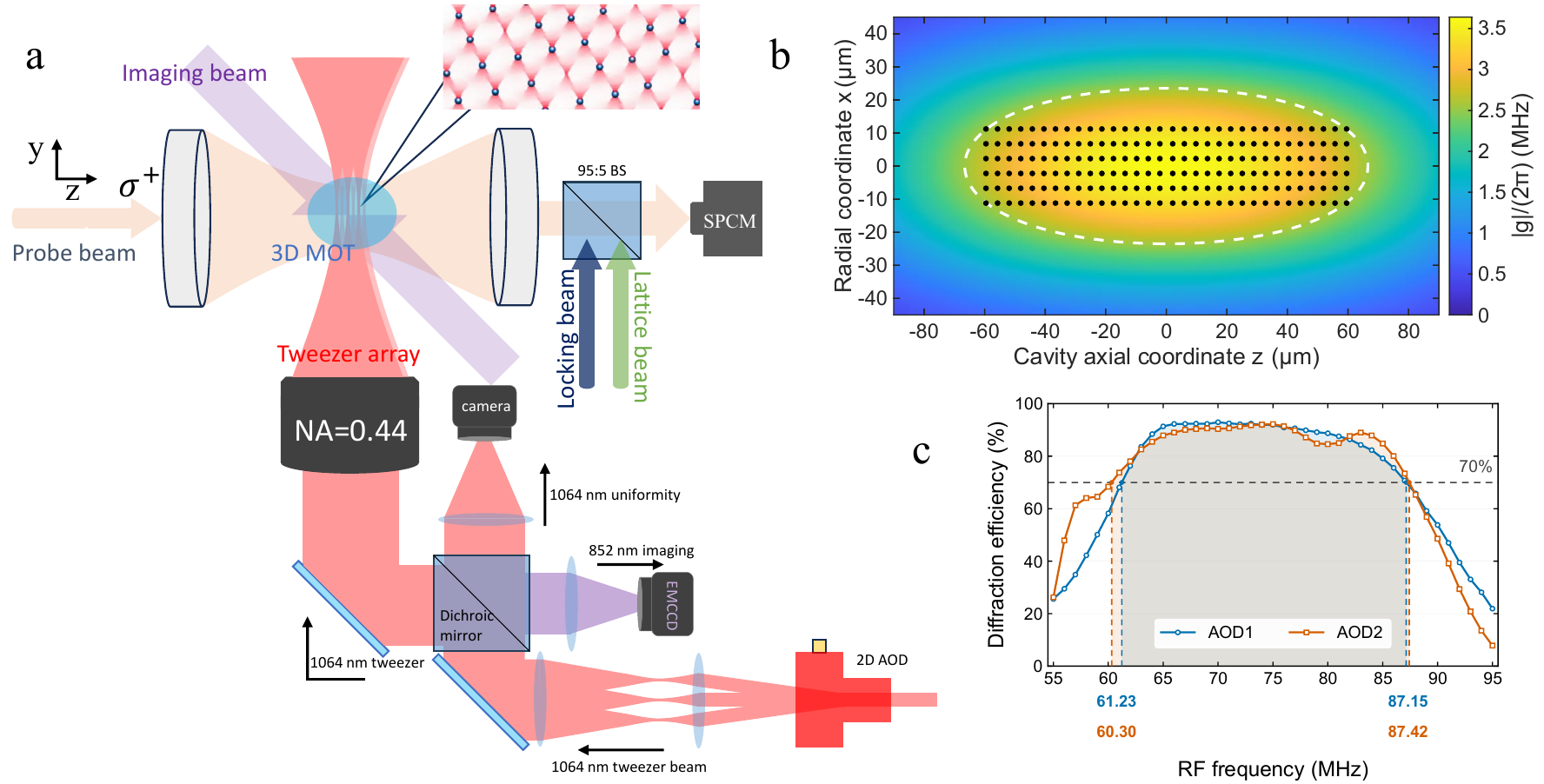}
  \caption{Experimental setup and array geometry. (a) Two-dimensional AOD-generated tweezer array
  inside the optical cavity. AOD1 and AOD2 control the radial and axial directions, respectively.
  The probe, cavity-locking, blue-detuned lattice, tweezer, and imaging beams and the fluorescence
  collection path are indicated. (b) Calculated coupling amplitude envelope sampled at the lattice
  nodes, with the 70\% contour and the sites of a \(32\times6\) array. (c) Measured diffraction
  efficiencies of the two AODs, with the RF intervals over which the efficiencies exceed 70\%
  indicated.}
  \label{fig:setup}
\end{figure*}

\section{Experimental system and array geometry}
\label{sec:system}

\subsection{Intracavity tweezer array and optical detection}

We generate a two-dimensional array of 1064 nm optical tweezers with two crossed acousto-optic
deflectors and focus it near the waist of the Fabry--Pérot cavity (Fig.~\ref{fig:setup}(a)). The
tweezers confine individual cesium atoms, while a blue-detuned intracavity lattice tightens axial
confinement near antinodes of the probe mode. We measure the collective optical response through
cavity probe transmission and identify the occupied traps by site-resolved fluorescence imaging.

The 1.315-mm-long cavity is formed by two mirrors, each with a radius of curvature of 50 mm. At 852
nm, the mirror transmissions are 0.23 ppm and 76.02 ppm, and the cavity finesse is about \(8.1
\times 10^{4}\). The fundamental transverse mode has a waist radius of \(39.4\,\mu\mathrm{m}\). For
the cesium cycling transition \(|\left. \ g \right\rangle \equiv 6S_{1/2}|\left. \ F = 4,\ m_{F} =
4 \right\rangle \leftrightarrow |\left. \ e \right\rangle \equiv 6P_{3/2}|\left. \ F = 5,\ m_{F} =
5 \right\rangle\), the cavity QED parameters are \(\left( \gamma,\kappa,g_{0} \right) = 2\pi \times
(2.617,\ 0.7,\ 3.64)\,\mathrm{MHz}\), where \(\gamma\) is the atomic polarization decay rate,
\(\kappa\) is the cavity field amplitude decay rate, and \(g_{0}\) is the theoretical maximum
single-atom coupling strength. The corresponding maximum single-atom cooperativity is \(C =
\frac{g_{0}^{2}}{2\kappa\gamma} = 3.6\).

A 1064 nm trapping beam passes through a fiber-coupled acousto-optic modulator and two crossed
AODs. Additional relay optics direct the diffracted beams into the vacuum cell from below,
perpendicular to the cavity axis. An objective with a numerical aperture of 0.44 focuses the beams
near the cavity center to form the tweezer array. We define z along the cavity axis and x
transverse to it in the array plane. AOD1 sets the trap positions along x, and AOD2 sets them along
z. The relay optics give approximately the same displacement per unit radio-frequency shift along
both axes. Both AODs are mounted on rotation stages to align the array with the cavity mode. Atomic
light-shift and trap oscillation measurements give a tweezer waist radius of approximately 1.33
\(\mu\mathrm{m}\).

For cavity spectroscopy, a weak probe enters through the lower-transmission mirror. A single-photon
counting module detects the light transmitted through the higher-transmission mirror, and a time
tagger records the photon numbers and arrival times. A locking beam at 847.7 nm stabilizes the
cavity length using the Pound--Drever--Hall technique, with the probe mode tuned to the cesium
cycling transition near 852.356 nm. A blue-detuned beam near 851 nm forms an intracavity lattice
that tightens axial confinement and helps localize the atoms near probe-mode
antinodes~\cite{liu2023coupling}. The locking and blue-detuned lattice beams enter through the
higher-transmission mirror via a beam splitter.

\subsection{Cavity coupling profile}

The tweezer array is centered near the cavity waist, and its axial extent is much shorter than the
Rayleigh range. We therefore neglect the axial variation of the Gaussian mode envelope. In the
array plane (y = 0), the position-dependent single-atom coupling to the probe mode is

\begin{equation}
  g(x,z)\  = \ g_{\max}\exp\!\left[ - \frac{x^{2}}{{w_{c}}^{2}}\right]\cos(k_{c}z)
  \label{eq:local-coupling}
\end{equation}

Here, \(w_{c}\) denotes the radius of the fundamental cavity mode, and \(k_{c} = 2\pi/\lambda_{c}\)
is the probe wavenumber. We take \(z = 0\) at a probe antinode at the cavity center.
Equation~\eqref{eq:local-coupling} describes the ideal spatial dependence. The effective coupling
can also depend on atomic motion, internal-state preparation, and probe polarization. We set
\(g_{\max}\) equal to the theoretical maximum \(g_{0}\) for the cycling transition specified above.

The blue-detuned intracavity lattice confines atoms near its intensity nodes. In the model, a
lattice node coincides with the probe antinode at the cavity center. Because the lattice and probe
wavelengths differ, the lattice nodes gradually shift relative to the nearest probe antinodes with
increasing distance from the center. Sampling Eq.~\eqref{eq:local-coupling} at the lattice nodes
gives a slowly varying envelope for the coupling magnitude,

\begin{equation}
  g_{\mathrm{env}}(x,z)\  = \ g_{\max}\exp\left\lbrack - \frac{x^{2}}{{w_{c}}^{2}} \right\rbrack\left| \cos(\Delta k\ z) \right|.
  \label{eq:envelope}
\end{equation}

Here, \(\Delta k\  = \ k_{L}\  - \ k_{c}\) is the wave-vector difference between the lattice and
probe fields. The Gaussian factor sets the transverse variation, while the difference in wavenumber
sets the axial variation of the envelope.

For the coupling map in Fig.~\ref{fig:setup}(b), we assign each axial tweezer center to the
nearest lattice node, with lattice period \(a_L=\lambda_L/2\). In the experiment, the tweezers
remain on, and the atomic positions and localization widths depend on the combined trapping
potential and the atomic temperature.

\subsection{Selection of array geometry and RF spacings}

We select the drive bands using the measured diffraction efficiencies of the two AODs
[Fig.~\ref{fig:setup}(c)] and the available trapping power. After two successive first-order
diffractions, the RF shifts add, giving the optical frequency at site \((m,n)\),
\begin{equation}
  \nu_{mn} = \nu_{00} + m\delta f_z + n\delta f_x,
  \label{eq:site-frequency}
\end{equation}
where \(\nu_{00}\) is the reference-site frequency and \(\delta f_z\) and \(\delta f_x\) are
the signed RF increments along the axial and radial directions, respectively. Equal spacing
magnitudes can give different sites the same optical frequency. If the beams overlap and share a
polarization component, their interference can persist under time
averaging~\cite{decruyenaere2026projections}. In preliminary tests, we observed spot distortions
with small, equal RF spacings. We seek well-formed spots with distinct optical frequencies and
check the beat notes against the motional resonances. Where beams overlap, beating can drive
motion through position modulation near a trap oscillation frequency or cause parametric heating
through depth modulation near twice that frequency~\cite{savard1997heating}.

The cavity coupling envelope then constrains the array footprint. The measured tweezer
displacement per unit RF shift is approximately \(4.274\,\mu\mathrm{m}/\mathrm{MHz}\) along both
axes. The calculated coupling envelope extends farther axially than radially
[Fig.~\ref{fig:setup}(b)]. We use its 70\% amplitude contour, corresponding to about half the
maximum single-atom cooperativity, to guide the placement of an array elongated along the cavity
axis. Accommodating many sites in this region requires small spacings compatible with stable trapping.

Together, these constraints lead to a \(32\times6\) array with axial and radial RF spacings of
\(0.90\,\mathrm{MHz}\) and \(1.05\,\mathrm{MHz}\), respectively. The outermost
tweezer centers span \(119.2\,\mu\mathrm{m}\) axially and \(22.4\,\mu\mathrm{m}\) radially.
For the nominal axial spacing of \(3.85\,\mu\mathrm{m}\), the modeled separations vary slightly
around nine lattice periods. The axial tones span
60.05--87.95~MHz and the radial tones span 71.875--77.125~MHz, with most tones falling in the
bands where the diffraction efficiency exceeds 70\%.

We take the site indices to increase with their respective optical frequency shifts. For the
chosen spacings, Eq.~\eqref{eq:site-frequency} gives the condition for exact degeneracy,
\begin{equation}
  6\Delta m + 7\Delta n = 0
  \quad\Longleftrightarrow\quad
  (\Delta m,\Delta n) = (7q,-6q),
  \label{eq:degeneracy}
\end{equation}
where \(q\) is an integer. With six radial sites, \(|\Delta n|\leq 5\) restricts the solution to
\(q=0\), so all intended beams have distinct optical frequencies. The minimum beat frequency is
\(0.15\,\mathrm{MHz}\), above the estimated parametric resonance at twice the trap oscillation
frequency (approximately 100~kHz).

\begin{figure*}[!tbp]
  \centering
  \includegraphics[width=0.91\textwidth]{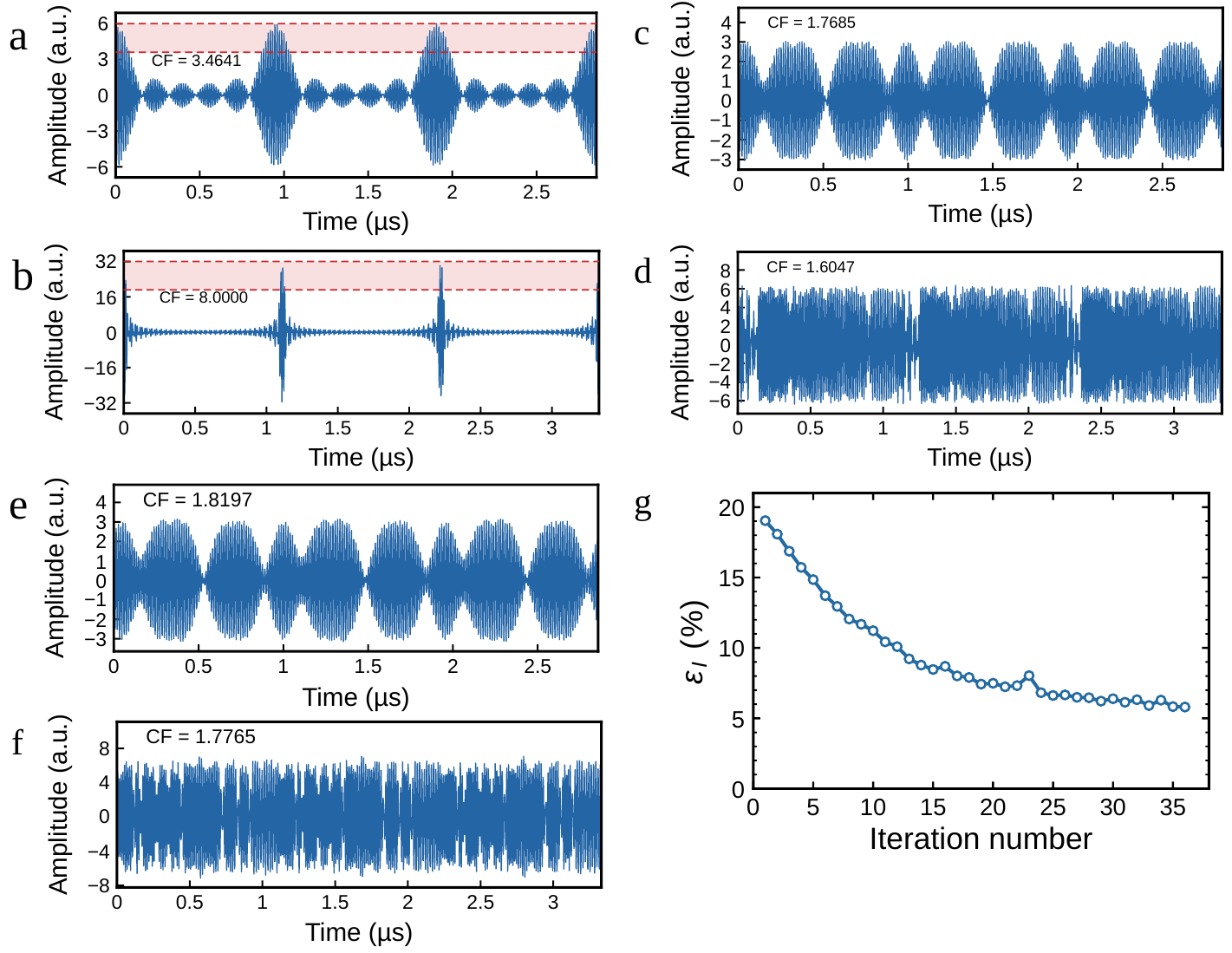}
  \caption{Multitone RF waveforms and intensity optimization. (a) and (b) Initial waveforms for the
  radial 6-tone and axial 32-tone channels, respectively, with equal relative amplitudes and zero
  relative phases. The red regions highlight large instantaneous peaks. (c) and (d) Corresponding
  waveforms after phase optimization at equal amplitudes. (e) and (f) Final waveforms after
  amplitude refinement using the optical intensity distribution and atomic fluorescence. All
  waveforms are calculated from the tone parameters programmed into the AWG. (g) Optical intensity
  nonuniformity during feedback optimization.}
  \label{fig:waveforms}
\end{figure*}

\section{Array preparation and calibration}
\label{sec:array}

\subsection{Multitone drive optimization and intensity uniformity}

The RF drive determines both the optical power delivered to the array and the uniformity of the
traps. For either AOD, the programmed waveform is a sum of tones,

\begin{equation}
  s(t)\  = \ \sum_{j = 1}^{N}{A_{j}\cos\left( 2\pi f_{j}t + \varphi_{j} \right)}
  \label{eq:rf-drive}
\end{equation}

where \(A_{j}\), \(f_{j}\), and \(\varphi_{j}\) are the voltage amplitude, frequency, and phase of
tone \(j\), and \(N\) is the number of tones. Large instantaneous peaks can drive the RF power
amplifier into its nonlinear gain region. This compression and nonlinear acousto-optic
responses can distort spot intensities, reduce usable drive power, and generate parasitic
diffracted beams~\cite{wang2023intermodulation,sun2024optimization,mittenbuehler2025synthesis}.

Following the joint phase and amplitude optimization approach of Ref.~\cite{sun2024optimization},
we quantify the maximum absolute voltage relative to the root-mean-square voltage by the crest
factor

\begin{equation}
  J\ \left( \bm{\varphi} \right)\  = \ \frac{\max_{t}\left| s(t) \right|}{\sqrt{\left\langle s^{2}(t) \right\rangle_{t}}}\ .
  \label{eq:crest-factor}
\end{equation}

Changing the relative phases redistributes the instantaneous peaks of the waveform while leaving
its root-mean-square voltage approximately unchanged. Reducing the crest factor allows the
available output range of the AWG to be used more effectively and reduces the risk of driving the
RF amplifier into a nonlinear regime.

To optimize the phases numerically, we normalize the sampled waveform by its root-mean-square
voltage to obtain \(u_{n}\) and approximate the maximum absolute sample value with the smooth
objective

\begin{equation}
  F_{\beta}\left( \bm{\varphi} \right)\  = \ \frac{1}{\beta}\ln\left\lbrack \frac{1}{N_{t}}\sum_{n = 1}^{N_{t}}{\exp\left( \beta\sqrt{{u_{n}}^{2} + \varepsilon^{2}} \right)} \right\rbrack
  \label{eq:phase-objective}
\end{equation}

where \(N_{t}\) is the number of optimization samples, \(\beta\) controls the approximation to the
maximum, and \(\varepsilon\) smooths the absolute value. We use \(\beta = 30\) and \(\varepsilon =
10^{- 6}\) and optimize all the phases jointly using analytical gradients.

For each AOD, we run the optimization from six initial phase sets, including one Schroeder-type
set, one Newman-type set, and four random sets. Each run ends at convergence or at the iteration
limit. We select the candidate with the lowest sampled crest factor \(J\) and reevaluate it over
the full arbitrary waveform generator (AWG) record.

At equal tone amplitudes and zero relative phases, the RF tones add constructively at certain
times, producing large peaks in the programmed waveforms (Fig.~\ref{fig:waveforms}(a) and
\ref{fig:waveforms}(b)). We optimized the relative phases while holding the tone amplitudes fixed.
The calculated crest factor, defined as the ratio of the peak absolute voltage to the
root-mean-square voltage, decreased from 3.4641 to 1.7685 for the six-tone radial drive and from
8.0 to 1.6047 for the 32-tone axial drive (Fig.~\ref{fig:waveforms}(c) and \ref{fig:waveforms}(d)).
The lower crest factor increases the usable output range of the AWG and reduces the risk of RF
amplifier compression.

We then hold the phases fixed while adjusting the RF amplitudes using the measured spot
intensities. With the camera operating in its linear, unsaturated regime, we obtain an intensity
measure \(I_{i}\) for each site by integrating a region of interest (ROI) and subtracting the local
background using a common procedure. Each RF tone affects a full row or column, so we evaluate the
response across the entire array after each adjustment. We quantify the relative spread of the
integrated intensities as

\begin{equation}
  \varepsilon_{I}\  = \ \frac{\operatorname{std}_{i}\left( I_{i} \right)}{\operatorname{mean}_{i}\left( I_{i} \right)}
  \label{eq:intensity-spread}
\end{equation}

We also inspect the spot shapes and widths and check for parasitic light, since the integrated
intensities alone provide an incomplete measure of trap depth.

We recalculate \(J\) after each amplitude update. If the crest factor increases appreciably or the
waveform exceeds the preset peak limit at the required output power, we reoptimize the phases at
the new amplitudes and remeasure the intensity distribution. We stop when both the intensity
criterion \(\varepsilon_{I} < \varepsilon_{\mathrm{set}}\) and the RF peak limit are satisfied.

Finally, we refine the RF amplitudes using site-resolved atomic fluorescence under fixed loading,
cooling, and imaging conditions. The fluorescence provides feedback on the combined effects of trap
depth, light shifts, and cooling at the atoms. We select settings that favor trapping and retention
and recalculate the crest factor after substantial amplitude updates.

Camera-based feedback adjusts the RF tone amplitudes to compensate for the frequency-dependent
optical response of the crossed deflectors. The relative standard deviation of the measured spot
intensities reaches the 6\% target after 36 iterations (Fig.~\ref{fig:waveforms}(g)). After further
amplitude refinement using atomic fluorescence, the final radial and axial crest factors are 1.8197
and 1.7765, respectively (Fig.~\ref{fig:waveforms}(e) and \ref{fig:waveforms}(f)), both within the
operating limits.

\begin{figure*}[!tbp]
  \centering
  \includegraphics[width=0.94\textwidth]{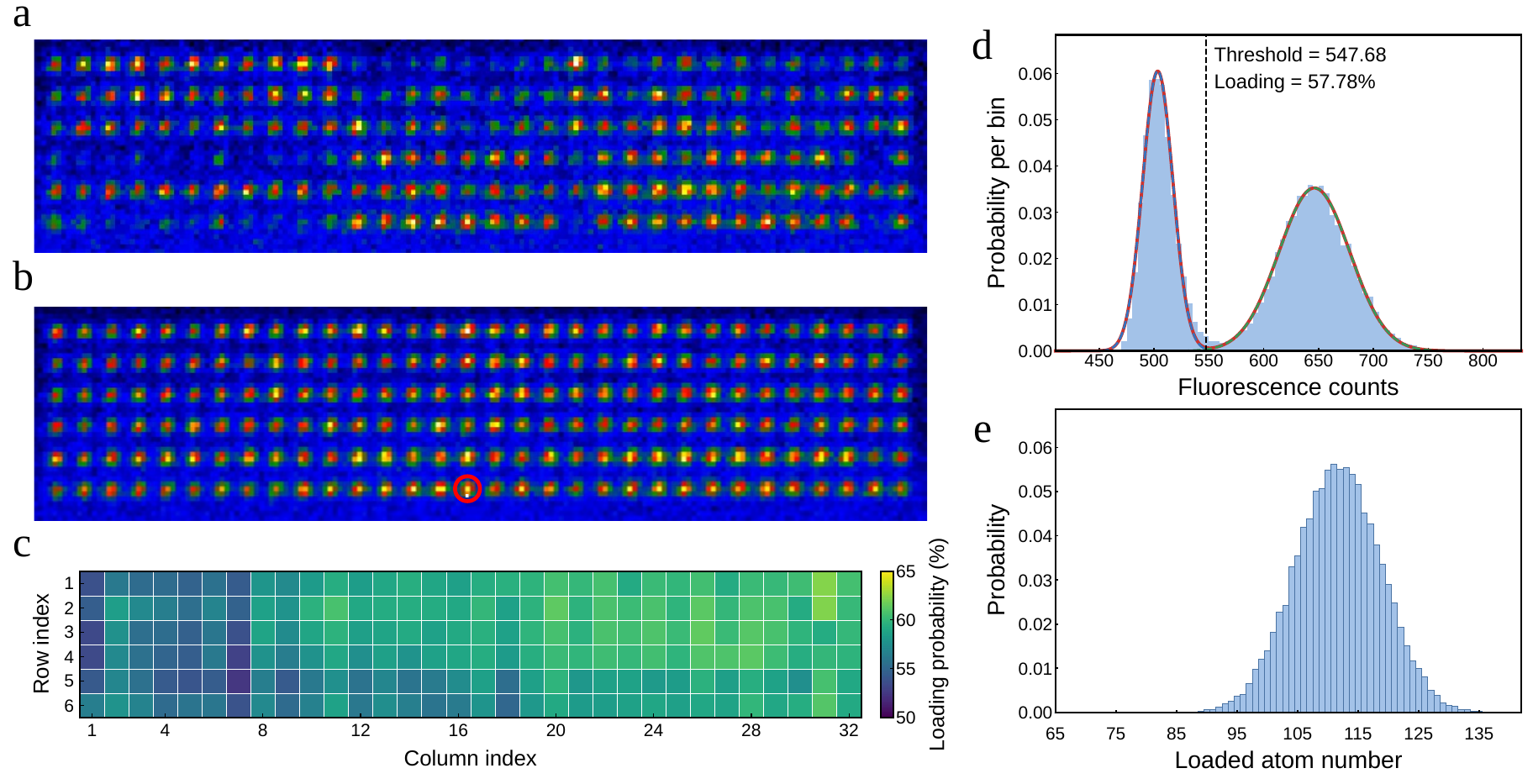}
  \caption{Atomic loading performance. Fluorescence images accumulated over several tens of
  independent loading trials after (a) optical intensity optimization and (b) further refinement
  using atomic fluorescence. Panels (c--e) summarize 30\thinspace{}000 loading trials. (c)
  site-resolved loading probabilities, (d) fluorescence histogram for the site in row 6, column 16,
  marked by the red circle in (b). The dashed line indicates the classification threshold. (e)
  distribution of the imaged atom number.}
  \label{fig:loading}
\end{figure*}

\subsection{Atom loading and site-resolved detection}

During loading, the diffraction efficiency under multitone drive is approximately 70\%. An average
optical power of about 5.6 mW per tweezer gives \(\frac{U}{k_{B}} \approx 600\,\mu\mathrm{K}\),
where \(U\) is the trap depth.

A three-dimensional magneto-optical trap (3D MOT) prepares a cold cesium cloud near the cavity
center for loading into the tweezers. Red-detuned light-assisted collisions suppress multiple
occupancy, and polarization-gradient cooling reduces the temperature of the trapped atoms. For
site-resolved detection, counterpropagating imaging beams with lin \(\perp\) lin polarizations
excite atomic fluorescence. The same objective used to form the tweezers collects the fluorescence
and images it onto an electron-multiplying charge-coupled device camera.

Figures~\ref{fig:loading}(a) and \ref{fig:loading}(b) compare fluorescence accumulated over
independent loading trials after camera-based intensity optimization and after further refinement
using atomic fluorescence. Several sites give low signals after intensity optimization alone. With
atomic feedback, the relative spread of the background-subtracted site fluorescence decreases from
40.43\% to 15.58\%. The accumulated fluorescence depends on both the occupation probability and the
brightness of an occupied site.

To quantify stochastic loading rate, we fit the fluorescence histogram from the spectroscopy trials
for each site to determine a threshold that distinguishes an empty trap from one occupied by a
single atom. Figure~\ref{fig:loading}(d) shows the histogram and fit for the site in row 6, column
16. In each trial, we compare the fluorescence signal at each site with its threshold to identify
the occupied sites and count the total number of loaded atoms. For each site, the loading
probability is the fraction of trials in which it is classified as occupied.

The mean loading probability across the 192 sites is approximately 58\%
(Fig.~\ref{fig:loading}(c)). Over 30\thinspace{}000 trials, the atom number per trial has a mean of
111.66 and a standard deviation of 7.10 (Fig.~\ref{fig:loading}(e)).

\begin{figure*}[!tbp]
  \centering
  \includegraphics[width=0.88\textwidth]{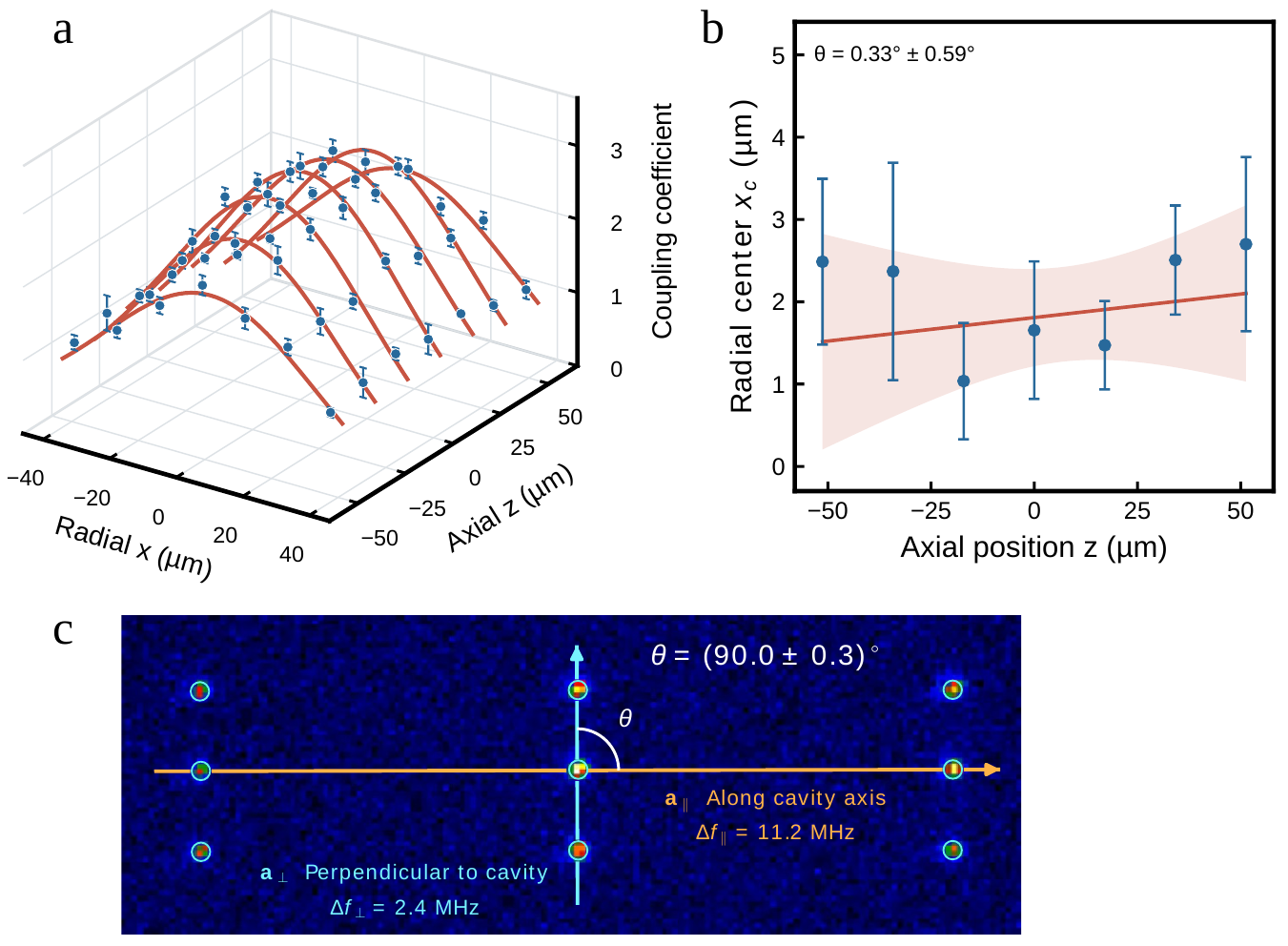}
  \caption{Spatial variation of single-atom coupling and array alignment. (a) Radial scans of the
  single-atom coupling in megahertz at several axial positions, with Gaussian fits. (b) Fitted
  radial centers \(x_{c}\) versus axial position \(z\). Shading indicates the 95\% confidence
  interval of the linear fit. (c) Site centers and fitted grid for a \(3\times3\) calibration array
  imaged by fluorescence. The axial and radial RF spacings are 11.2 and 2.4 MHz, corresponding to
  approximately 47.9 and 10.3 \(\mu\mathrm{m}\), respectively.}
  \label{fig:alignment}
\end{figure*}

\subsection{Spatial dependence of single-atom coupling and array alignment}

We align the array with the cavity mode using the two AOD rotation stages. To determine the
orientation of the axial array direction, we map the coupling of a single trapped atom to the
cavity. At each axial position \(z_{j}\), we scan the tweezer along the radial direction and
extract the coupling from the cavity transmission spectra (Fig.~\ref{fig:alignment}(a)). Treating
the atom as pointlike on the scale of the transverse mode profile, we fit each profile with

\begin{equation}
  g_{j}(x)\  = \ A_{j}\exp\lbrack - \frac{{(x\  - \ x_{c,j})}^{2}}{{w_{j}}^{2}}\rbrack
  \label{eq:radial-profile}
\end{equation}

where \(x_{c,j}\) is the radial center, \(A_{j}\ \) is the peak coupling strength, and \(w_{j}\) is
the radial distance from the center at which the fitted coupling falls to \(1/e\) of its peak. We
fit the radial centers as a function of axial position to determine the tilt of the axial array
direction relative to the cavity axis.

The scans cover most of the array's axial extent and extend across its radial span. At each axial
position, a Gaussian fit gives the radial center of the mode. The fitted peak coupling also varies
across the sampled axial positions.

The fitted centers are plotted in Fig.~\ref{fig:alignment}(b). A linear fit of radial centers to
\(x_{c}(z) = a + bz\) yields \(b = 0.0057 \pm 0.0103\), corresponding to an axial tilt \(\theta =
\arctan(b) = 0.33{^\circ} \pm 0.59{^\circ}\). Over the \(119.2\,\mu\mathrm{m}\) axial span of the
array, the fitted slope corresponds to a change \(|b|L_{z} = 0.68\,\mu\mathrm{m}\) in the radial
mode center, about 1.7\% of the cavity waist radius. The fitted tilt is consistent with zero within
its uncertainty.

We adjust AOD1 so that the radial array direction is orthogonal to the axial direction set by AOD2.
To measure this angle, we extract the site centers of a \(3\times3\) calibration array from
fluorescence images and fit them to an affine grid,

\begin{equation}
  r_{mn}\  = \ r_{0}\  + \ mu\  + \ nv,\ \ \ \ \alpha\  = \ \cos^{- 1}\left( \frac{u\  \cdot \ v}{|u||v|} \right)
  \label{eq:grid}
\end{equation}

where the two basis vectors describe the axial and radial directions of the fitted grid, and
\(\alpha\) is the angle between them.

The grid fit to the site positions in a separate \(3\times3\) calibration array gives an angle of
\((90.0 \pm 0.3){^\circ}\) between the two array axes (Fig.~\ref{fig:alignment}(c)). The two
calibrations show that the long axis is aligned with the cavity axis and that the array axes are
orthogonal within the measurement uncertainties.

\begin{figure*}[!tbp]
  \centering
  \includegraphics[width=0.82\textwidth]{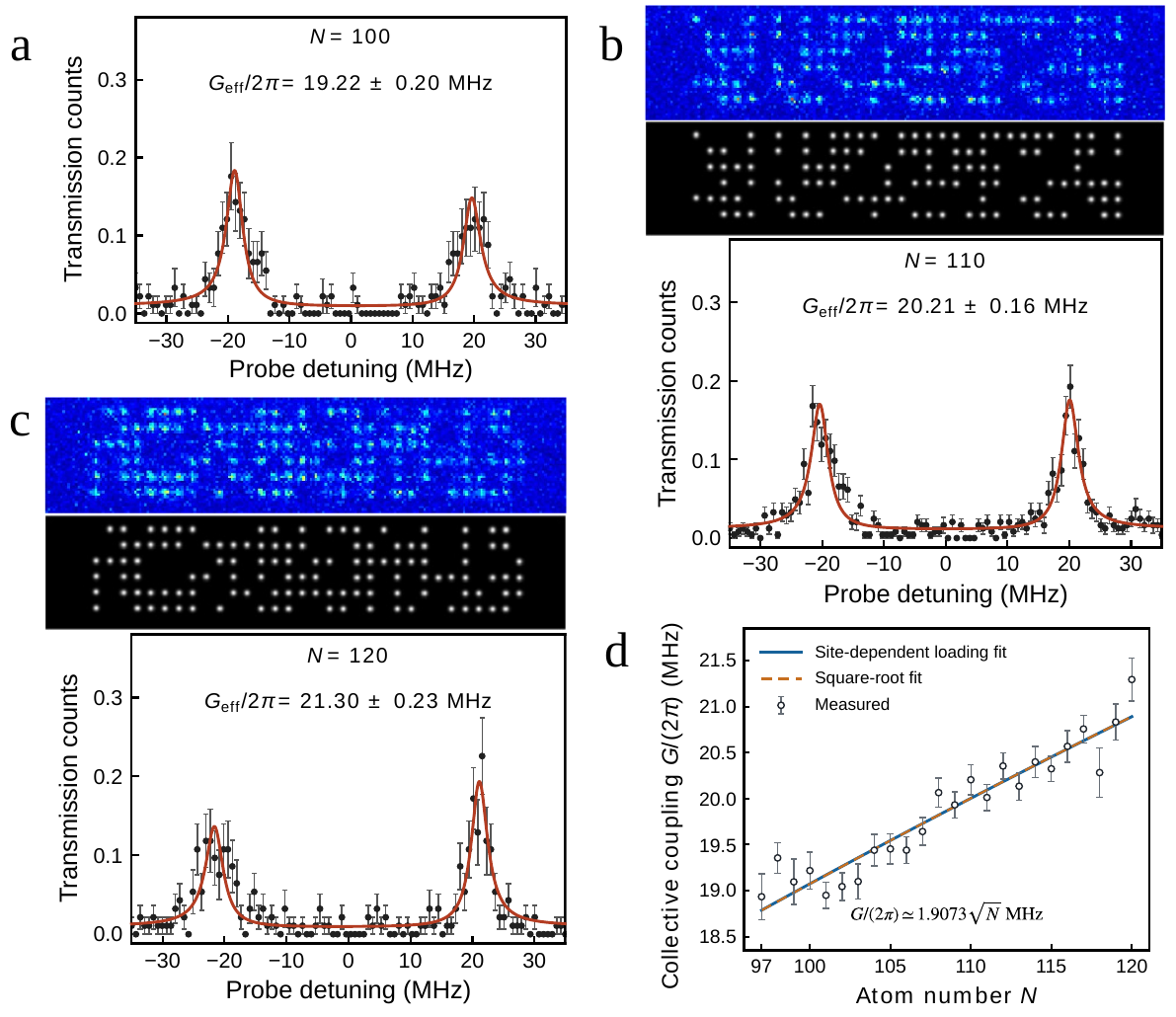}
  \caption{Collective coupling to a single cavity mode. (a--c) Mean transmission spectra and fits
  for groups with 100, 110, and 120 imaged atoms, respectively. The images in (b) and (c) show
  representative single-shot fluorescence images and the corresponding grayscale reconstruction of
  the identified site occupations. (d) Effective collective coupling strength
  \(G_{\mathrm{eff}}/2\pi\) as a function of imaged atom number N. The solid blue curve is a fit
  based on the calculated coupling envelope and measured site-resolved loading probabilities. The
  dashed orange curve is a square-root fit.}
  \label{fig:collective}
\end{figure*}

\section{Collective atom--cavity coupling}
\label{sec:collective}

\subsection{Cavity transmission spectra}

For spectroscopy, we reduce the average tweezer depth from 600 \(\mu\mathrm{K}\) to approximately
200 \(\mu\mathrm{K}\) and increase the depth of the blue-detuned lattice to improve axial
localization near the probe-mode antinodes.

In the weak-excitation regime, we describe the cavity transmission using the linear response of the
Tavis--Cummings model~\cite{tavis1968hamiltonian}. We define the probe--cavity and probe--atom
detunings as \(\Delta_{c} = \omega_{p} - \omega_{c}\) and \(\Delta_{a} = \omega_{p} - \omega_{a}\),
respectively, where the subscripts p, c, and a denote the probe, empty cavity, and atomic
resonance. For a collective coupling strength \(G\), the transmission normalized to the
empty-cavity peak is

\begin{equation}
  T(\omega_{p})\  = \ \frac{\kappa^{2}}{\left|\kappa\  - \ i\Delta_{c}\  + \ \frac{G^{2}}{\gamma\  - \ i\Delta_{a}}\right|^{2}}
  \label{eq:transmission}
\end{equation}

We group the cavity transmission spectra by the number of atoms identified in the fluorescence
images. The mean spectra for groups with 100, 110, and 120 imaged atoms show resolved normal-mode
doublets (Fig.~\ref{fig:collective}(a)--\ref{fig:collective}(c)). Fits to the weak-probe
transmission model in Eq.~\eqref{eq:transmission} give effective collective couplings
\(G_{\mathrm{eff}}/2\pi\) of \(19.22 \pm 0.20\), \(20.21 \pm 0.16\), and \(21.30 \pm 0.23\) MHz,
respectively. Figures~\ref{fig:collective}(b) and \ref{fig:collective}(c) also show representative
single-shot fluorescence images and the corresponding occupation patterns. Over the full analyzed
range of 97--120 imaged atoms, \(G_{\mathrm{eff}}/2\pi\) ranges from 18.9 to 21.3 MHz.

\subsection{Spatially weighted coupling model}

Because the single-atom coupling varies with position (Fig.~\ref{fig:alignment}(a)), configurations
with equal atom numbers can have different collective couplings. To estimate how spatial weighting
affects the atom-number dependence, we combine the calculated coupling envelope
(Fig.~\ref{fig:setup}(b)) with the measured site loading probabilities (Fig.~\ref{fig:loading}(c)).
The model uses occupation probabilities conditioned on total atom number.

For a particular configuration of occupied sites, the collective coupling is the quadrature sum of
the single-atom couplings,

\begin{equation}
  G_{\mathrm{eff}} = \sqrt{\sum_{i = 1}^{M}n_{i}\left| g_{i} \right|^{2}}
  \label{eq:collective-coupling}
\end{equation}

where \(M\) is the total number of sites, \(n_{i} \in \{ 0,1\}\) is the occupation of site \(i\),
and \(g_{i}\) is its single-atom coupling strength. For the idealized spatial profile used in the
theoretical comparison,

\begin{equation}
  g_{i}^{\mathrm{th}} = g_{0}h_{i},\quad\quad h_{i} = e^{- x_{i}^{2}/w^{2}}\left| \cos\left( \Delta kz_{i} \right) \right|
  \label{eq:site-weight}
\end{equation}

where \(h_{i}\) is the spatial coupling factor at site \(i\) and \(w\) is the cavity waist radius
used in Eq.~\eqref{eq:envelope}. At a fixed total atom number \(N\), \(q_{i}(N)\) is the
conditional probability that the site is occupied, with \(\sum_{i}q_{i}(N) = N\). The
root-mean-square collective coupling predicted by the model is

\begin{equation}
  \begin{aligned}
    G_{\mathrm{th,rms}}(N) &\equiv \sqrt{\langle G^{2}\rangle_{N}} = g_{0}\sqrt{N\mathcal{A}_{N}}, \\
    \mathcal{A}_{N} &= \frac{1}{N}\sum_{i}q_{i}(N)\left| h_{i} \right|^{2}
  \end{aligned}
  \label{eq:weighted-coupling}
\end{equation}

We model site occupations as independent Bernoulli variables with the measured loading
probabilities as their parameters. At fixed N, the conditional occupation probability of a site is
its measured loading probability multiplied by the probability of one fewer occupied site among the
remaining sites and divided by the probability of that total among all sites. The spatially
weighted fit uses the atom-number dependence in Eq.~\eqref{eq:weighted-coupling} with an overall
fitted scale. For the ideal reference, we evaluate Eq.~\eqref{eq:weighted-coupling} with the
theoretical maximum single-atom coupling.

The root-mean-square coupling of the configurations at a given atom number provides a reference for
the scaling. The effective coupling obtained from the averaged spectrum can differ from that
root-mean-square value because the transmission in Eq.~\eqref{eq:transmission} is a nonlinear
function of the coupling.

Figure~\ref{fig:collective}(d) compares the spatially weighted fit with a fit proportional to
\(\sqrt{N}\). The two curves have nearly the same shape over the measured range. The square-root
fit gives \(G_{\mathrm{eff}}/(2\pi) = (1.9073 \pm 0.0032)\sqrt{N}\,\mathrm{MHz}\). Using the
theoretical maximum single-atom coupling \(g_{0,\mathrm{th}}/(2\pi) = 3.64\,\mathrm{MHz}\), the
spatial model gives an ideal reference coefficient \(g_{0,\mathrm{eff}}/(2\pi) =
g_{0,\mathrm{th}}\sqrt{\mathcal{A}_{N}}/(2\pi) = 3.21\,\mathrm{MHz}\). The measured coefficient is
approximately 59.3\% of this reference value.

\section{Discussion and conclusion}
\label{sec:discussion}

Under the independent-loading model, the mean squared spatial coupling factor \(\mathcal{A}_{N}\)
in Eq.~\eqref{eq:weighted-coupling} varies from 0.77984 to 0.78010 for groups with 97--120 imaged
atoms, a relative change of approximately \(3.3\times10^{-4}\). The model therefore predicts that
the mean squared collective coupling grows nearly linearly with atom number even though the
single-atom coupling varies across the array. Its root-mean-square prediction and the square-root
fit have nearly identical shapes over this interval (Fig.~\ref{fig:collective}(d)). At the
precision of the measurement, the spectra are consistent with both trends.

The measured scaling coefficient is approximately 59\% of the ideal spatially weighted reference.
Finite axial motion, atom loss between imaging and probing, and imperfect internal-state
preparation may contribute to the discrepancy. Measuring atom survival across the spectroscopy
sequence and calibrating single-atom couplings under the same conditions would help identify the
dominant contributions. More effective axial cooling could increase the coupling by localizing the
atoms closer to the probe-mode antinodes. Generating the array with a static hologram on a spatial
light modulator would remove the beat notes associated with the differential optical frequency
shifts of the crossed deflectors and provide a way to assess their contribution to heating.

In conclusion, we extend collective strong coupling to the hundred-atom scale in a two-dimensional
array with stochastic loading and site-resolved occupation detection. The combination of
site-resolved detection and collective strong coupling provides a basis for using a single optical
mode to probe and manipulate a large neutral-atom array. For quantum simulation, suitable optical
drives could turn the shared mode into a mediator of long-range spin exchange. State-resolved
imaging could then reveal how correlations develop across the array under the combined effects of
photon-mediated interactions and cavity loss. For quantum computing, state-selective coupling could
enable cavity-assisted readout of selected atoms within a register. With the addition of coherent
atom--photon interfaces at selected sites, the same architecture could provide optical links
between registers, extending the role of the cavity from collective operations within a processor
to quantum communication between processors.

\section*{Acknowledgements}

This work was supported by Innovation Program for Quantum Science and Technology (Grant No. 2023ZD0300400); National Key Research and Development Program of China (2021YFA1402002); National Natural Science Foundation of China (NSFC) (Grants No. 92465201, 12474360, and 92265108).

%\section*{Author Contributions}

%%T.-C.Z. and G.L. supervised the project. S.-J.G., Z.-H.W., G.L., and T.-C.Z. designed the experiment. S.-J.G., G.-S.T., Z.-H.W., P.-F.Y., and P.-F.Z. performed the experiment. S.-J.G., Z.-G.H., Z.-H.W., G.L., and T.-C.Z. analyzed the data and wrote the paper. All the authors discussed the results and contributed to the writing of the paper. 

%\section*{Data availability}

%All data supporting the findings of this study are available within the article and its
%Supplementary Information.

%\section*{Code availability}

%Available when needed.

%\section*{Conflict of Interest}

%The authors declare no competing interests.

\bibliography{references}

\end{document}